\documentclass[twocolumn,floatfix,amsmath,amssymb,superscriptaddress]{revtex4}
\usepackage[dvips]{graphicx}
\usepackage{amsfonts}
\usepackage{dcolumn}
\usepackage{bm}
\usepackage{color}
\usepackage{epsfig}
\usepackage{hyperref}
\usepackage{bbold}

\newcommand{\kett}[1]{\left| #1 \right\rangle}

\newcommand{\be}{\begin{equation}}
\newcommand{\ee}{\end{equation}}
\newcommand{\bea}{\begin{eqnarray}}
\newcommand{\eea}{\end{eqnarray}}

\def\g{\gamma}
\def\G{\Gamma}
\def\d{\delta}

\def\e{\epsilon}

\def\th{\theta}

\def\L{\Lambda}
\def\m{\mu}

\def\p{\pi}

\def\r{\rho}

\def\vf{\varphi}

\def\w{\omega}

\def\blr{{\mathbf r}}

\def\callI{\mbox{$\mathcal{I}$}}
\def\callJ{\mbox{$\mathcal{J}$}}

\def\iif{\infty}

\def\1op{\hat{\mathbbm{1}}}
\def\1{\mathbbm{1}}
\def\nn{\nonumber}

\begin{document}

\title{Benchmarking Nonequilibrium Green's Functions 
against Configuration Interaction for time-dependent 
Auger decay processes}

\author{F. Covito}
\affiliation{Max Planck Institute for the Structure and Dynamics of 
Matter and Center for Free-Electron Laser  Science, Luruper Chaussee 
149, 22761 Hamburg, Germany}  
\author{E. Perfetto}
\affiliation{CNR-ISM, Division of Ultrafast Processes in Materials 
(FLASHit), Area della ricerca di Roma 1, Monterotondo Scalo, Italy}
\affiliation{Dipartimento di Fisica, 
Universit\`a di Roma Tor Vergata, Via della Ricerca Scientifica,
00133 Rome, Italy}
\author{A. Rubio}
\affiliation{Max Planck Institute for the Structure and Dynamics of 
Matter and Center for Free-Electron Laser  Science, Luruper Chaussee 
149, 22761 Hamburg, Germany}  
\affiliation{Center for Computational Quantum Physics (CCQ), The 
Flatiron Institute, 162 Fifth avenue, New York NY 10010} 
\affiliation{Nano-Bio Spectroscopy Group, Universidad del Pa\'i's
Vasco, 
20018 San Sebasti‡n, Spain}  
\author{G. Stefanucci}
\affiliation{Dipartimento di Fisica, Universit\`a di Roma Tor
Vergata, Via della Ricerca Scientifica, 00133 Rome, Italy}
\affiliation{INFN, Sezione di Roma Tor Vergata, Via della Ricerca
Scientifica 1, 00133 Roma, Italy}
\date{\today}

\begin{abstract}
We have recently proposed a Nonequilibrium Green's Function (NEGF) approach 
to include Auger decay processes in the ultrafast 
charge dynamics of photoionized molecules. 
 Within the so called Generalized 
Kadanoff-Baym Ansatz the fundamental unknowns of the NEGF equations are the reduced 
one-particle density matrix of bound electrons and the occupations of the continuum 
states. Both unknowns are one-time functions like the density in Time-Dependent 
Functional Theory (TDDFT). In this work we assess the accuracy of the 
approach against Configuration Interaction (CI) calculations in  
one-dimensional model systems. Our results show that NEGF correctly 
captures qualitative and quantitative features of the relaxation 
dynamics provided that the 
energy of the Auger electron is much larger than the Coulomb 
repulsion between two holes in the valence shells. 
For the accuracy of the results 
dynamical electron-electron correlations or, equivalently, memory 
effects play a pivotal role. The combination of 
our NEGF approach with the Sham-Schl\"uter equation may provide 
useful insights for the development of TDDFT exchange-correlation potentials with 
a history dependence. 

\end{abstract}
\maketitle

\section{Introduction}
Photo-ionized many-body systems relax to lower energy states
through nuclear rearrangement and charge redistribution. 
Nuclear dynamics does typically play a role on longer time 
scales, although there are  situations where 
electron-nuclear and electron-electron interactions compete on the 
same timescale, e.g., in the vicinity of a conical intersection.
% The timescale of the relaxation processes depends on the energy 
% excited states involved. 
% take place on different timescales althogh . While
% nuclear rearrangements happen on the nanosecond to microsecond
% timescale \textcolor{cyan}{(please check this)}, electron dynamics
% arise much faster, down to attoseconds. The timescale of the
% relaxation processes depends directly on the energy of the out of
% equilibrium state. 
At the (sub)femtosecond timescale, however, the most relevant 
relaxation channel of core-ionized molecules
is the Auger decay which is  exclusively driven by the Coulomb 
interaction~\cite{Pazourek-RevModPhys.87.765}. 

Recent advances in pump-probe experiments 
made it possible to follow the attosecond dynamics of atoms 
after the sudden expulsion of a core 
electron~\cite{uiberacker2007attosecond,Uphues2008,drescher2002time,zherebtsov2011attosecond,Schins-PhysRevLett.73.2180}. 
Theoretical frameworks describing the Auger decay have been proposed,
the more accurate being the ones based on many-body 
wavefunctions, see also Ref.~\cite{Kazansky.2011}.
Although these methods are in principle applicable to atoms as well 
as molecules,
they quickly become prohibitive for systems with more than a few
active electrons. For instance, Auger decays in ionized small 
molecules or molecules of biological interest are 
extremely difficult to cope with wavefunction approaches due to 
the large number of states involved in the process.
Still, Auger decays contribute to the 
relaxation dynamics of these more complex systems, which are 
currently attracting an increasing interest and 
attention~\cite{RevModPhys.81.163,GI.2014,SLMPS.2016,RTT.2017,Nisoli-review}.
It is therefore crucial to develop first-principles approaches
capable of capturing the (sub)femtosecond relaxation mechanisms
induced by electronic correlations and applicable to atoms as well as 
molecules.

The most widely used method for large scale real-time simulations is 
Time-Dependent Density Functional 
Theory~\cite{RungeGross:84,Ullrich:12,Maitra.2016} 
(TDDFT), which gives an
adequate and computationally affordable tool for the description of
systems consisting of up to thousands of atoms. The most efficient and
extensively used functionals for TDDFT calculations are the 
space-time local exchange-correlation (xc) functionals. 
It has been shown numerically in
Ref.~\cite{PhysRevB.86.045114} that these approximate functionals fail in capturing Auger
decays, the fundamental reason being that they lack memory effects -- the xc potential 
depends on the instantaneous density only.

We have recently proposed a first-principles  NonEquilibrium Green's Function (NEGF)
approach~\cite{Covito2018} which overcomes the limitation of 
adiabatic functionals
and that may inspire new ideas for the inclusion of memory effects
in the TDDFT functionals. The method is applicable to molecules with 
up to tens of atoms and at its core there is an equation to 
simulate the electron dynamics 
in the parent cation without dealing explicitly with 
the Auger electrons. The idea is similar in spirit to
the embedding scheme in 
time-dependent quantum transport where the electron dynamics in 
the  molecular junction is simulated without dealing explicitly with 
the electrons in the 
leads~\cite{ksarg.2005,vsa.2006,Stefanuccipumping,mssvl.2009,spc.2010}. However, whereas in quantum transport  
the integration out of electrons in the leads gives an embedding
self-energy which is independent of the density in the junction, 
the integration out of the Auger electrons gives an Auger self-energy 
which is a functional of the density in the molecule. 

In order to assess the quality of the  NEGF approach in this 
work we use the time-dependent charge distribution of the bound 
electrons to reconstruct the Auger wavepacket in free space, and then benchmark the 
results against exact configuration interaction (CI) calculations. 
We perform NEGF and CI simulations in 
a model one-dimensional (1D) system and study the real space-time shape
of the Auger wavepacket as well as the Auger spectrum. The main outcome 
of this investigation is that the results of the NEGF approach are in excellent 
agreement with those from CI  provided that the repulsion 
between the valence holes is much smaller than the energy of the 
Auger electron.

\section{Description of the system and theory}
\label{systemsec}
Let us consider a 1D finite system described by the one-particle 
Hartree-Fock (HF) basis
$\{\varphi_{i},\vf_{\m}\}$, where roman indices run over bound states 
and greek indices run over continuum states.
The equilibrium
Hamiltonian can be conveniently written as the sum of 
three terms
\begin{equation}
	\hat{H}^{\rm eq}=\hat{H}_{\rm bound}+\hat{H}_{\rm
Auger}+\hat{H}_{\rm 
	cont},
	\label{eqham}
\end{equation}
where $\hat{H}_{\rm bound}$ is the bound electrons Hamiltonian,
$\hat{H}_{\rm Auger}$ is the Auger interaction and $\hat{H}_{\rm
cont}$ is the free-continuum part. In our basis, these are written as
\begin{subequations}
	\begin{align}
		&\hat{H}_{\rm bound} = \sum_{ij}h_{ij}\hat{c}^{\dag}_{i}\hat{c}_{j}
		+\frac{1}{2}\sum_{ijmn}v_{ijmn}
		\hat{c}^{\dag}_{i}\hat{c}^{\dag}_{j}\hat{c}_{m}\hat{c}_{n},\\
		&\hat{H}_{\rm Auger}=\sum_{ijm}\sum_{\mu}
		\left(v^{A}_{ijm\mu}\hat{c}^{\dag}_{i}\hat{c}^{\dag}_{j}\hat{c}_{m}\hat{c}_{\mu}
		+{\rm H.c.}\right),\\
		&\hat{H}_{\rm
cont}=\sum_{\mu}\epsilon_{\mu}\hat{c}^{\dag}_{\mu}\hat{c}_{\mu},
	\end{align}
\end{subequations}
where $c^\dag_i$ ($c_i$) is the creation (annihilation) operator for
the state $\vf_{i}$ (the same convention applies to the continuum 
index $\mu$), $h_{ij}$ are the one-electron integrals,  $\epsilon_{\mu}$
are the continuum single-particle HF energies 
and $v_{ijmn}$
($v^{A}_{ijm\mu}$) are the two-electron Coulomb integrals 
responsible for intra-molecular (Auger)
scatterings. The one- and two-electron integrals are defined as
\begin{subequations}
	\begin{align}
 		h_{ij} &\equiv \int dx \varphi^\star_i(x)
[-\frac{1}{2}\nabla^2_x+V_n(x)]\varphi_j(x),\\
		v_{ijmn} &\equiv \int dx dx'\varphi^\star_i(x) \varphi^\star_j(x')
V_{e}(x,x') \varphi_m(x')\varphi_n(x), \label{twobody_int}
	\end{align}
\end{subequations}
with $V_n(x)$ and $V_e(x,x')$ the nuclear and electron-electron
potential. Note that the Auger Coulomb integrals $v^{A}_{ijm\mu}$ are
defined according to Eq.~\eqref{twobody_int} with $n=\mu$. In
Eq.~\eqref{eqham} we discard all the off-diagonal
contribution 
$h_{i\mu}$, $h_{\mu\mu'}$ as well as all Coulomb integrals with more than one
index in the continuum. This approximation does not
affect the physical description of the dynamics as demonstated 
by comparisons against full grid calculations in 
Ref.~\cite{Covito2018}. In fact, in the HF basis both $h_{i\mu}$ and 
$h_{\mu\mu'}$ are much smaller than $h_{ij}$ and $\e_{\m}$ whereas 
Coulomb integrals with two or more indices in the continuum are 
responsible for scattering process that are highly suppressed by 
phase-space arguments if the photoelectron energy is much 
larger than the kinetic energy of the Auger electron. Henceforth, this 
condition is assumed to be fulfilled.

The explicit simulation of the ionization process with a
laser field does not represent a complication for the 
NEGF method. In fact, the general framework presented in  Ref.~\cite{Covito2018}
accounts for the coupling of external fields with 
the bound-bound and bound-continuum dipole matrix elements. Instead, 
the framework 
discards the coupling of external fields with the continuum-continuum dipole matrix 
elements and, therefore, light-field streaking experiments
relevant to, e.g., attosecond metrology~\cite{attometrology}, or 
multuphoton ionization processes are left out.

In this work we focus  on the dynamics induced by the sudden 
removal of a core electron, thus the ionization process is not 
simulated.
An additional simplification used for the simulations below (which is 
however not 
essential for the approach) consists in keeping 
only integrals of the form $v^A_{c\mu v_1v_2}$, where
$c$ labels the state of the suddenly created core hole, $v_1$ and $v_2$ 
label two valence states and $\m$ an arbitrary
continuum state. We also observe that the HF wavefunctions are real 
since the Hamiltonian is invariant under time-reversal. This implies 
that the Coulomb integrals have the following symmetries
\be
v_{ijmn}=v_{jinm}=v_{imjn}=v_{njmi}
\ee
and the like with $n\to\m$.

\subsection{NEGF equations}
The derivation of the NEGF equations within the so called Generalized 
Kadanoff-Baym Ansatz~\cite{PhysRevB.34.6933} (GKBA) has been presented 
elsewhere~\cite{Covito2018}; 
here we only describe the structure of these equations without entering into the 
complex mathematical and numerical details.

Let $\r$ be the one-particle reduced density matrix in the bound 
sector and $f_{\m}$ be the occupations of the continuum states. Then 
the NEGF equations read
\be
\left\{
\begin{array}{l}
\dot{\r}=-i\left[h_{\rm HF}[\r],\r\right]
-\callI[\r,f]-\callI^{\dag}[\r,f]
\\ \\
\dot{f}_{\m}=-\callJ_{\m}[\r,f]-\callJ^{\ast}_{\m}[\r,f]
\end{array}
\right.,
\label{CHEERSeq}
\ee
where the single-particle HF Hamiltonian is defined according to
\be
h_{{\rm HF},ij}=h_{ij}+\sum_{mn}(v_{imnj}-v_{imjn})\r_{nm}.
\ee
The matrix $\callI$ and the scalar $\callJ_{\m}$ at time $t$ are 
explicit functionals of $\r$ and $f$ at all previous times. They
are evaluated using the so-called second-Born (2B)
approximation which has been shown to contain the fundamental 
scattering of the Auger  
process~\cite{PhysRevB.39.3489,PhysRevB.39.3503}. 
The dependence on $\r$ and $f$ occurs through the lesser and greater 
GKBA Green's functions~\cite{PhysRevB.34.6933} 
\be
G^{\lessgtr}(t,\bar{t})=\mp
\left[G^{\rm R}(t,t')\r^{\lessgtr}(t')-\r^{\lessgtr}(t)G^{\rm A}(t,t')\right],
\ee
and the like for $G^{\lessgtr}$ with indices in the continuum. Here,
the retarded ($G^{\rm R}$) and advanced ($G^{\rm A}$)
Green's functions are evaluated in 
the HF approximation (and hence they are functionals of $\r$ and $f$ 
too). The functional $\callI$  ($\callJ_{\m}$)
is linear in $G^{\lessgtr}$ with indices in the continuum  and 
quartic (cubic) in $G^{\lessgtr}$ with indices in the bound sector.
Their 
calculation requires to perform an integral from some initial time, say $t=0$, up 
to time $t$. The implementation of Eqs.~\eqref{CHEERSeq} does 
therefore scale
quadratically with the number of time steps. 
Notice that by setting $\callI=\callJ_{\m}=0$ is equivalent to perform time-dependent HF 
simulations. Like the adiabatic approximations in TDDFT, HF is local 
in time and therefore it is unable to describe Auger decays.

The scaling of the calculation of $\callI$ and $\callJ_{\m}$ 
with the number of basis functions is 
$\max[(N_{\rm bound})^{\mathfrak{p}},(N_{\rm bound})^{\mathfrak{q}}N_{\rm 
cont}]$, where 
 $N_{\rm bound}$ is the number of bound states, 
$N_{\rm cont}$ the number of continuum states and the exponents $3\leq 
\mathfrak{p}\leq 5$, 
$2\leq\mathfrak{q}\leq 4$ depend on the number of nonvanishing 
Coulomb integrals~\cite{Covito2018}. 
Currently, both $\callI$ and  
$\callJ_{\m}$ are implemented in the CHEERS code~\cite{PS-cheers} 
which, for $\callJ_{\m}=0$, has been recently used to study the charge 
transfer dynamics in a  donor-C$_{60}$ model dyad~\cite{C60paper2018} and  
the ultrafast charge migration 
in the phenylalanine aminoacid up to 40~fs~\cite{PSMS.2018}.
Since the calculation of $\callJ_{\m}$ is not heavier than the 
calculation of $\callI$,
the NEGF approach can be used to study time-dependent Auger processes 
driven by XUV or X-ray pulses 
in molecules with up to tens of atoms.

\subsection{CI calculation} 
Let us consider the simplest possible 
case of a system with one occupied core state, one occupied valence state
and a continuum of empty states.  We are interested in describing
the evolution of
the system starting from the initial state
\begin{equation}
	\kett{\phi_x} =
c^\dag_{c\uparrow}c^\dag_{v\downarrow}c^\dag_{v\uparrow} \kett{0},
\end{equation}
representing a core-hole of down spin. The evolution operator defined 
by the Hamiltonian in Eq.~(\ref{eqham}) mixes $\kett{\phi_x}$ with 
(we recall that only Coulomb integrals of the form $v_{c\m vv}$ and 
the like related by symmetries are nonvanishing, see Section~\ref{systemsec})
\begin{subequations}
	\begin{align}
		\kett{\phi_g} =
c^\dag_{c\uparrow}c^\dag_{c\downarrow}c^\dag_{v\uparrow} \kett{0},\\
		\kett{\phi_\m} =
c^\dag_{c\uparrow}c^\dag_{c\downarrow}c^\dag_{\m\uparrow} \kett{0},
	\end{align}
\end{subequations}
where $\kett{\phi_g}$ is the ``intermediate'' state with the filled
core, i.e., the ground state of the parent cation, and 
$\kett{\phi_\m}$ is the state describing the dication with an Auger 
electron in the continuum state $\m$. Carrying out the calculations 
it is easy to show that these states are coupled by the Hamitonian as 
follows 
\begin{subequations}
	\begin{align}
		& \hat{H}^{\rm eq} \kett{\phi_x} = E_x \kett{\phi_x} + T \kett{\phi_g}
+ \sum\nolimits_\m V_\m \kett{\phi_\m},\\
		& \hat{H}^{\rm eq} \kett{\phi_g} = E_g \kett{\phi_g} + T
\kett{\phi_x},\\
		& \hat{H}^{\rm eq} \kett{\phi_\m} = E_\m 
		\kett{\phi_x} + V_\m
\kett{\phi_x},
	\end{align}
	\label{Haction}
\end{subequations}
where the energies $E_x$, $E_g$, $E_\m$, $T$ and $V_\m$ are
given by 
\begin{subequations}
	\begin{align}\label{en_scales_def}
		&E_x = h_{cc} + 2h_{vv} + 2 v_{cvvc}+v_{vvvv} - v_{cvcv},\\
		&E_g = 2h_{cc} + h_{vv} + 2v_{cvvc} +v_{cccc} - v_{cvcv},\\
		&E_\m = 2h_{cc} + \e_{\m} + v_{cccc},\\
		&T = h_{cv} + v_{ccvc} + v_{cvvv},\\
		&V_\m = v_{vvc\m}.
	\end{align} 
	\label{energyscales}
\end{subequations}
The simplification brought about by the HF basis is now
evident. The HF Hamiltonian $h_{{\rm HF},ij} = h_{ij} +
\sum_k^{\rm occ}(2v_{i k k j} - v_{i k j k} )$ is diagonal
in the HF basis, therefore
\begin{equation}
	\begin{split}
		0 = h_{{\rm HF},cv} = & h_{cv} + (2v_{cccv}-v_{ccvc}) +
(2v_{cvvv}-v_{cvvv}) \\
		= & h_{cv} + v_{cccv}+v_{cvvv} \equiv T.
	\end{split}
\end{equation}
Thus the ``intermediate'' state $\kett{\phi_g}$ decouples from the
dynamics. 

We write the three-body wave function at time $t$ as 
\begin{equation}
	\kett{\psi(t)} = a_x(t)\kett{\phi_x} + \sum_\m a_\m(t) 
	\kett{\phi_\m}, 
\end{equation}
with initial condition $\kett{\psi(0)}=\kett{\phi_x}$. Taking into 
account Eqs.~(\ref{Haction}), the time-dependent Schr\"odinger 
equation yields a set of coupled equations for the coefficients of 
the expansion
\begin{equation}\label{CIeq}
	\begin{cases}
		i \dot{a}_x(t) = E_x a_x(t) + \sum\nolimits_\m V_\m 
		a_\m(t)\\
		i \dot{a}_\m (t) = V_\m a_x(t) + E_\m a_\m (t)
	\end{cases}
\end{equation}
to be solved with boundary conditions $a_x(0)=1$ and $a_k(0)=0$.

From the definitions in Eqs.~(\ref{energyscales}) 
it follows that for the continuum three-body state to have the same energy 
of the initial state, i.e., $E_{\m}=E_x$, the energy $\e_{\m}$ of the Auger electron 
has to be
\begin{equation}\label{aug_en_CI}
	\e_{\m}\equiv
	\epsilon^{\rm CI}_{\rm Auger} = 2 \epsilon_v^{\rm HF}
- \epsilon_c^{\rm HF} - v_{vvvv},
\end{equation}
where 
\begin{subequations}
	\begin{align}\label{en_scales_def}
		&\epsilon_c^{\rm HF}=h_{cc}+v_{cccc}+2v_{cvvc}-v_{cvcv},\\
		&\epsilon_v^{\rm 
		HF}=h_{vv}+v_{vvvv}+2v_{vccv}+v_{vcvc},
	\end{align} 
	\label{HFenergies}
\end{subequations}
are the core
and valence HF energies, respectively. It is therefore reasonable to 
expect a peak in the continuum occupations $f_{\m}$ for the $\m$ 
corresponding to an energy close to the value in Eq.~(\ref{aug_en_CI}).

In the next Section we solve numerically Eqs.~(\ref{CIeq}). However, 
in order to get some physical insight into the solution we here make a 
``wide-band-limit approximation'' (WBLA) and carry on the analytic treatment a 
bit further. Integrating the second equation (\ref{CIeq}) we have
\begin{equation}
a_{\m}(t)=-i\int_{0}^{t}dt'e^{-iE_{\m}(t-t')}V_{\m}a_{x}(t'),
\label{amuax}
\end{equation}
which correctly satisfies the boundary conditions $a_{\m}(0)=0$. 
Substituing this result into the first equation (\ref{CIeq}) we get
\be
i\dot{a}_x(t)=E_{x}a_{x}(t)+\int_{0}^{\infty}dt'K(t-t')a_{x}(t'),
\label{axclosed}
\ee
where
\begin{align}
K(t-t')&=-i\th(t-t')\sum_{\m}V_{\m}^{2}e^{-iE_{\m}(t-t')}
\nn\\
&\equiv\int\frac{d\w}{2\p}e^{-i\w(t-t')}\left[\L(\w)-\frac{i}{2}\G(\w)\right],
\label{kernel}
\end{align}
and
\be
\L(\w)-\frac{i}{2}\G(\w)=\sum_{\m}\frac{V_{\m}^{2}}{\w-E_{\m}+i0^{+}}.
\label{sigauger}
\ee
The real function $\L$ is connected to  $\G$ through a Hilbert 
transform, i.e.,
\be
\L(\w)=\int\frac{d\w'}{2\p}\frac{\G(\w')}{\w-\w'}\;,
\label{hilbert}
\ee
and from Eq.~(\ref{sigauger}) it is easy to show that
\begin{equation}
	\Gamma(\omega) = 2 \pi \sum\nolimits_\m V_\m^2 \delta(\omega 
	-E_\m).
	\label{gammadef}
\end{equation}

For systems in a box of lenght $L$ the continuum wavefunctions are 
proportional to $1/\sqrt{L}$ and hence $V^{2}_{\m}$ scales like 
$1/L$, see definition in Eq.~(\ref{twobody_int}). In the limit 
$L\to\iif$ the discrete sum in Eq.~(\ref{gammadef}) becomes an 
integral and $\Gamma(\omega)$ becomes a smooth function of $\w$. 
Assuming that $E_{x}$ is a few 
times  larger than $\Gamma(E_{x})$ and that $\Gamma(\omega)$ is a slowly 
varying function  for  $\omega\simeq
E_x$, we can then neglect the frequency 
dependence in $\G$: 
\begin{equation}
	\Gamma(\omega) \simeq \Gamma(E_x) \equiv \gamma,
	\label{wbla}
\end{equation}
which implies, see Eq.~(\ref{hilbert}), that we can approximate 
$\L\simeq 0$, see Eq.~(\ref{hilbert}). This is the so called WBLA, according to which the 
kernel $K$ in Eq.~(\ref{kernel}) can be approximated as
\be
K(t-t')=-\frac{i}{2}\g \,\d(t-t').
\ee
Substituing this result into Eq.~(\ref{axclosed}) and then using 
Eq.~(\ref{amuax}) it is  straighforward to find the following analytic 
solution
\begin{subequations}
	\begin{align}
		& a_x(t) = e^{-i E_x t - \frac{\gamma}{2} t},\\
		& a_\m(t) = - V_\m \frac{e^{-i (E_x - \frac{i}{2} \gamma)t}-e^{-i E_\m
t}}{E_\m - E_x + \frac{i}{2}\gamma}.
	\end{align}
	\label{analyticsol}
\end{subequations}
From Eqs.~(\ref{analyticsol}) we infer that the occupation of the continuum 
states is peaked at $E_{\m}=E_{x}$ or, equivalently, at 
$\e_{\m}=\epsilon^{\rm CI}_{\rm Auger}$,  in agreement with the discussion 
above Eq.~(\ref{aug_en_CI}). We emphasize that this conclusion is 
based on the WBLA. The exact solution contains a small correction 
which is proportional to the Hilbert transform of $\G(\w)$ at 
frequency $\w\simeq E_{x}$. 

\subsection{Comparing NEGF with CI}
\label{NEGFvsCI}

In the NEGF approach at the 2B level of approximation
two holes, in addition to feel an average 
(HF) potential generated by all other electrons, scatter directly 
 once. However, for a strong enough 
repulsion $v_{vvvv}$ it is necessary to include multiple 
valence-valence scatterings to predict the correct energy of the Auger 
electron. In fact, the red shift   
$v_{vvvv}$ in Eq.~(\ref{aug_en_CI}) can be captured only by 
summing multiple scatterings to infinite order (T-matrix 
approximation)~\cite{Sawatzky1977,CINI1977}.
Since the 2B approximation includes 
just a single scattering, the predicted Auger energy is
\begin{equation}\label{aug_en_2B}
	\epsilon_{\rm Auger}^{\rm 2B} = 2 \epsilon_v^{\rm HF} - \epsilon_c^{\rm HF}.
\end{equation}
In 3D molecules the neglect of $v_{vvvv}$
has only a minor impact on the internal (bound-electrons) dynamics 
since $v_{vvvv}$ is typically less than 1 eV and 
$\G(\w)$ varies rather slowly on this energy scales. 
In this work, however, we are 
also interested in the description of the Auger 
wavepacket. Taking into account that the 
repulsion $v_{vvvv}$ in 1D systems is larger than in 3D ones, 
a  sizable difference between the CI and 2B results
has to be expected. To demonstrate that such a
difference does not affect the overall physical picture nor
the details of the  Auger wavepacket but only the speed at which 
the Auger electron is expelled, we isolate the effects of 
multiple valence-valence scatterings from the CI formulation. 
Let us express the energy 
$E_x$ defined in Eq.~\eqref{en_scales_def} in terms of HF energies
\begin{equation}
	E_x= 2\epsilon_v^{\rm HF} + \epsilon_c^{\rm HF} - v_{cccc}
-4v_{vccv}+2v_{vcvc} - v_{vvvv}.
\label{newEx}
\end{equation}
The HF energy $\epsilon_v^{\rm HF}$ is blue shifted by $v_{vvvv}$, 
see Eq.~(\ref{HFenergies}), an 
effect captured by the 2B approximation. The effect of multiple 
scatterings manifests in the red shift given by the last term of 
Eq.~(\ref{newEx}).
In the next Section we show that solving 
Eqs.~\eqref{CIeq} using for $E_{x}$ the value in Eq.~(\ref{newEx}) 
with  $v_{vvvv}=0$ one recovers the NEGF results (notice that this is 
not equivalent to set $v_{vvvv}=0$ in the Hamiltonian since this 
Coulomb integral renormalizes the HF energy $\epsilon_v^{\rm HF}$).
We will refer to 
this CI approximation as CI2B.

\section{Results}
We consider a one-dimensional (1D) atom with
soft Coulomb interactions. This particular example is a  severe
test for the NEGF method since the continuum spectrum has a strong
frequency dependence and the valence-valence repulsion energy is
of the same order of magnitude of the Auger energy.

The 1D atom is defined on the points $x_n=na$ of a 1D grid, with
$|n|\le N_{\rm grid}/2$. In our model the Coulomb interaction is
different from zero only in a box of radius $R$ centered around the 
nucleus. The
one-body Hamiltonian on the grid reads
\be
h(x_{n},x_{m})=\delta_{n,m}[2\kappa+V_{n}(x_{n})]-\delta_{|n-m|,1}\kappa
\ee
with $V_n(x)=U_{\rm en}/\sqrt{x^{2}+a^{2}}$ the nuclear potential
and $\kappa$ the hopping integral between neighbouring points. 
Electrons interact through $v(x,x')=ZU_{\rm
ee}/\sqrt{(x-x')^{2}+a^{2}}$. We analyze the system using
$N_{\rm grid}=1601$ grid-points and choose the parameters according to
(atomic units are used throughout):
$a=0.5$,  $\kappa=2$, $Z=4$, $U_{\rm en}=2$, $U_{\rm ee}=U_{\rm
en}/2$ and $R=10a$. With four electrons 
the HF spectrum  has five
bound states (per spin), the lowest two of which are occupied. The
energies of the occupied levels are $\epsilon^{\rm HF}_c=-4.33$ and
$\epsilon^{\rm HF}_v=-1.65$ for the core and 
valence  respectively, yielding a 2B Auger energy $\e^{\rm 
2B}_{\rm Auger}=1.02$. We work in the sudden creation approximation, 
according to which
the system is perturbed by suddenly removing a core electron. 
In the NEGF approach this is simulated by subtracting to the 
equilibrium density matrix $\r^{\rm eq}_{ij}$ an 
infinitesimal amount of charge from the core, hence 
$\r_{ij}(0)=\r^{\rm eq}_{ij}-\d_{ic}\d_{jc}n_{h}$. In the results 
below the hole density $n_{h}=0.04$.

\begin{figure}[t]
	\centering
	\includegraphics[width=\columnwidth]{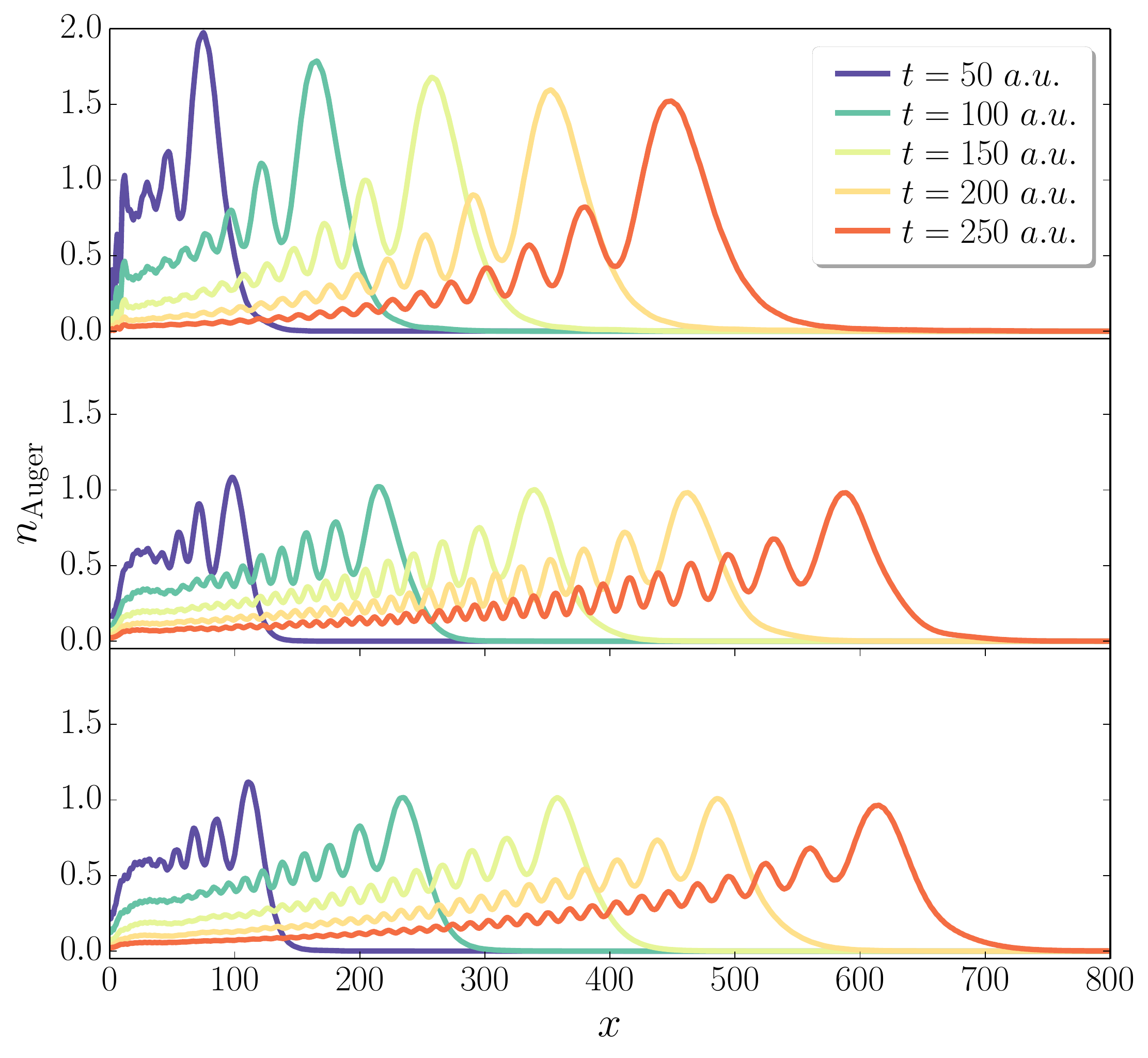}
	\caption{Snapshots of the density of the Auger wavepacket leaving
the atom (nucleus is situated in $x=0$) calculated using CI
(top),  NEGF approach (middle) and CI2B
 (bottom). The vertical axes have been rescaled by a factor
$10^{4}$ for all curves.}
	\label{fig:naug}
\end{figure}
Subsequently to the creation of the core hole, the Auger process
starts taking place, triggering an internal electron dynamics
(refilling of the core state) and the expulsion of charge toward the 
continuum states.   The
time-dependent  occupation of the core state $n_c(t)$ is
predicted in both CI and 2B calculations to have the following behavior $n_c(t)
= 1-n_he^{-\Gamma t}$, where $n_h$ is the core hole created and
$\Gamma$ is the inverse lifetime of the Auger decay. 
Due to the neglect of multiple scatterings,  the Auger 
decay is faster in 2B and the corresponding $\G$ is  overestimated by 
a factor 1.5. As already pointed out, this discrepancy is expected to be 
much smaller in 3D molecules since the
valence-valence repulsion is not as large.

In Fig. \ref{fig:naug} we display snapshots at different times 
of the real-space density of the Auger wavepacket
as obtained by performing 
 CI (top), NEGF (middle) and CI2B
calculations (bottom). The results
in the NEGF approach closely resemble the ones in the CI2B treatment, 
in agreement with the discussion in Section~\ref{NEGFvsCI}.
The CI calculation, as expected, shows a slower wavepacket. However, 
the overall shape, i.e., asymmetric packet with superimposed 
accumulating ripples on the tail, is common to all methods.
\begin{figure}[t]
	\centering
	\includegraphics[width=\columnwidth]{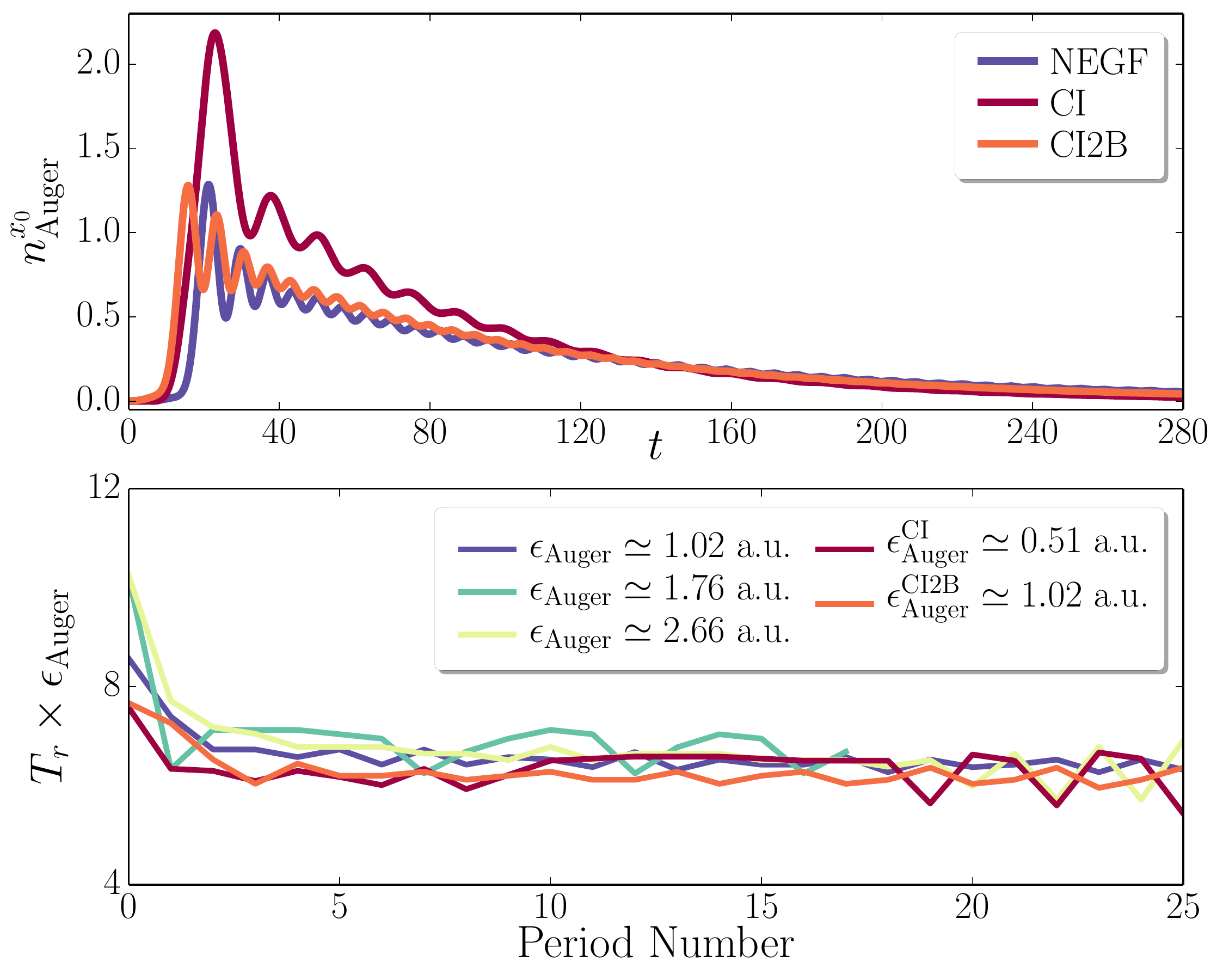}
	\caption{The top panel shows the time-dependent density of the Auger
wavepacket at a fixed distance $x_{0}=30$ from the nucleus  for 
NEGF, CI and CI2B. The bottom panel displays the
period of the ripples at $x_{0}$ versus the number of elapsing periods
for the three calculations of the top panel and for two more NEGF 
calculations, see main text.}
	\label{fig:ripples}
\end{figure}
We mention that the amplitude of the ripples as well as the wavefront 
of the Auger wavepacket change if, instead of the sudden creation of 
a core-hole, we 
would have simulated the ionization process using an external laser 
pulse. In fact, these features are not universal and depend  
on the intensity and duration of the 
perturbing field~\cite{Covito2018}. On the other hand, the time $T_r$ elapsing between two 
consecutive maxima at any fixed position 
is an intrinsic feature of the Auger decay, following the law
\begin{equation}\label{T_r}
	T_r = \frac{2 \pi}{\epsilon_{\rm Auger}}.
\end{equation}
In the top panel of  Fig.
\ref{fig:ripples} we show the time-dependent density $n_{\rm 
Auger}({x_0},t)$ of the Auger
wavepacket  at a certain distance $x_{0}$
from the
nucleus. The densities exhibit ripples of different frequency since 
the energy of the 
Auger electron is different  in CI, NEGF and CI2B. The small discrepancy between 
 NEGF and CI2B  is due to the fact that the solution in 
Eqs.~(\ref{analyticsol}) is valid only in the WBLA. Taking into 
account the frequency dependence of $\G$ one would find a small 
correction to $E_{\m}-E_{x}$ proportional to the Hilbert transform of 
$\G$. From the top panel of  Fig.~\ref{fig:ripples} we see that this 
correction is rather small and therefore the 
WBLA is an excellent approximation in this case.  

In the bottom panel of Fig.
\ref{fig:ripples} we show the value  of the time $T_r$ elapsing between two 
consecutive maxima of the wavepacket 
versus the number of maxima (counted starting from the left most 
maximum in the top panel). In the figure  $T_r$ is rescaled 
by the Auger energy. In all cases, after a short transient phase, 
$T_r$ attains the value $2\p$. In addition to the values of $T_{r}$ 
corresponding to the three curves of the top panel, 
in the bottom panel we also report 
the trend of $T_{r}$ calculated in Ref.~\cite{Covito2018}
for two  more   
NEGF simulations. 
More specifically, we considered 
two different combinations of range and
strengths of the Coulomb interactions $(R,U_{\rm en},U_{\rm
ee})=(100a,2.6,2.08),\;(10a,2.7,2.025)$ yielding Auger
electrons at energies $\epsilon^{\rm 2B}_{\rm Auger} = 1.76,~2.66$ 
respectively. As we can see, 
the quantity $T_r \times \epsilon_{\rm Auger}$ remains independent of 
the system.

\begin{figure}
	\centering
	\includegraphics[width=\columnwidth]{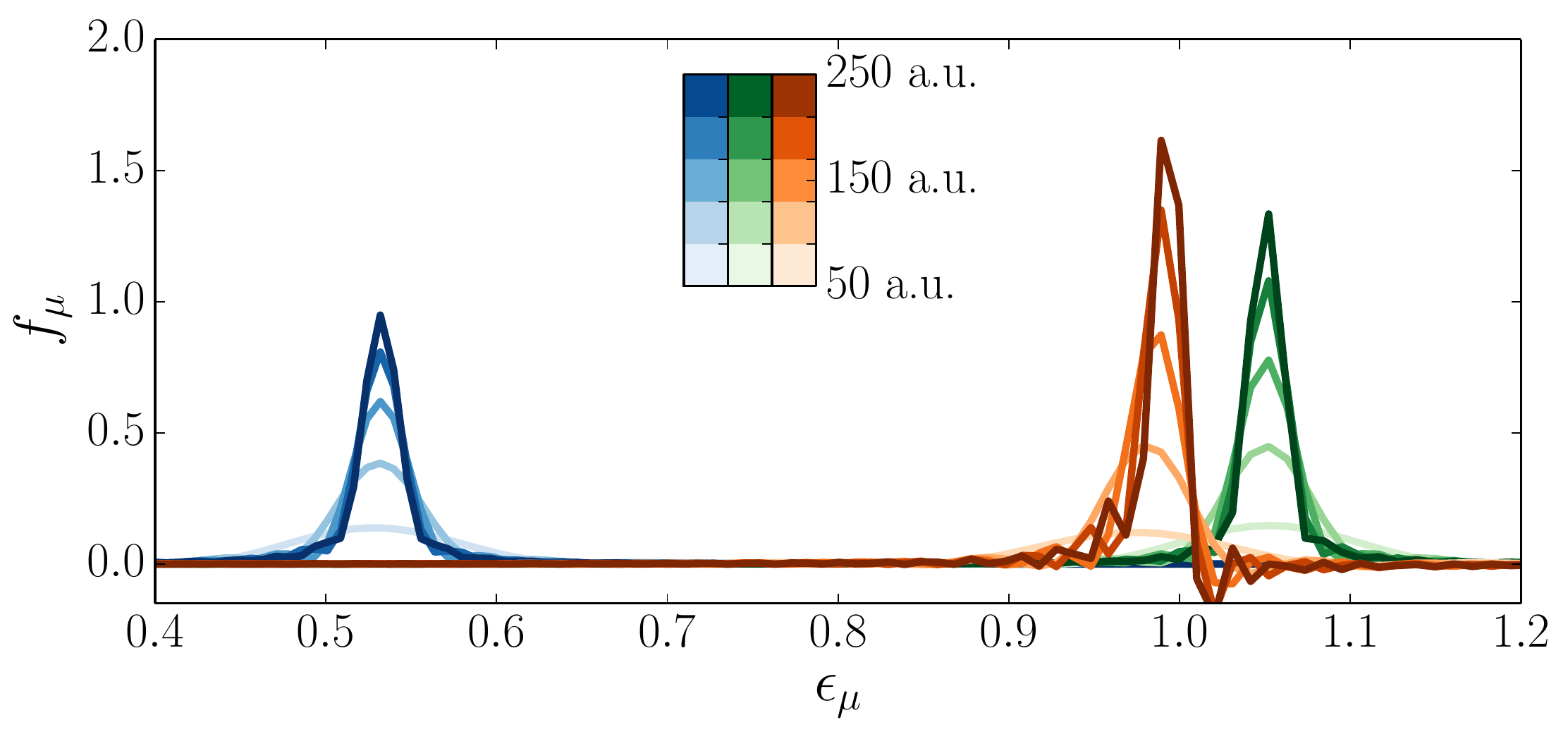}
	\caption{Snapshots of the occupations $f_\mu$ of the continuum
states versus their energy $\epsilon_\mu$ for 
CI (blue), CI2B (green) and NEGF (orange). The times of the snapshots 
(from light to dark) are given by the color bars.}
	\label{fig:fmu}
\end{figure}

Finally, in Fig. \ref{fig:fmu} we display the snapshots of the
time-dependent occupations $f_\mu(t)$ of the continuum states
$\varphi_\mu$. After
the creation of the core-hole, occurring at $t=0$, the continuum
states start to get populated and, as time passes, gradually 
get peaked around the Auger energy $\epsilon^{\rm
CI}_{\rm Auger} \simeq0.51$ for the CI calculation and
$\epsilon^{\rm 2B}_{\rm Auger}
\simeq1$ for the NEGF and CI2B calculation -- the
small deviation between these two calculations has been discussed 
previously.

\section{Conclusions}
%\bibliography{/Users/gianlucastefanucci/files/ARTICOLI/mybiblio}

\begin{thebibliography}{33}
\expandafter\ifx\csname natexlab\endcsname\relax\def\natexlab#1{#1}\fi
\expandafter\ifx\csname bibnamefont\endcsname\relax
  \def\bibnamefont#1{#1}\fi
\expandafter\ifx\csname bibfnamefont\endcsname\relax
  \def\bibfnamefont#1{#1}\fi
\expandafter\ifx\csname citenamefont\endcsname\relax
  \def\citenamefont#1{#1}\fi
\expandafter\ifx\csname url\endcsname\relax
  \def\url#1{\texttt{#1}}\fi
\expandafter\ifx\csname urlprefix\endcsname\relax\def\urlprefix{URL }\fi
\providecommand{\bibinfo}[2]{#2}
\providecommand{\eprint}[2][]{\url{#2}}

\bibitem[{\citenamefont{Pazourek et~al.}(2015)\citenamefont{Pazourek, Nagele,
  and Burgd\"orfer}}]{Pazourek-RevModPhys.87.765}
\bibinfo{author}{\bibfnamefont{R.}~\bibnamefont{Pazourek}},
  \bibinfo{author}{\bibfnamefont{S.}~\bibnamefont{Nagele}}, \bibnamefont{and}
  \bibinfo{author}{\bibfnamefont{J.}~\bibnamefont{Burgd\"orfer}},
  \bibinfo{journal}{Rev. Mod. Phys.} \textbf{\bibinfo{volume}{87}},
  \bibinfo{pages}{765} (\bibinfo{year}{2015}),
  \urlprefix\url{https://link.aps.org/doi/10.1103/RevModPhys.87.765}.

\bibitem[{\citenamefont{Uiberacker et~al.}(2007)\citenamefont{Uiberacker,
  Uphues, Schultze, Verhoef, Yakovlev, Kling, Rauschenberger, Kabachnik,
  Schr{\"o}der, Lezius et~al.}}]{uiberacker2007attosecond}
\bibinfo{author}{\bibfnamefont{M.}~\bibnamefont{Uiberacker}},
  \bibinfo{author}{\bibfnamefont{T.}~\bibnamefont{Uphues}},
  \bibinfo{author}{\bibfnamefont{M.}~\bibnamefont{Schultze}},
  \bibinfo{author}{\bibfnamefont{A.~J.} \bibnamefont{Verhoef}},
  \bibinfo{author}{\bibfnamefont{V.}~\bibnamefont{Yakovlev}},
  \bibinfo{author}{\bibfnamefont{M.~F.} \bibnamefont{Kling}},
  \bibinfo{author}{\bibfnamefont{J.}~\bibnamefont{Rauschenberger}},
  \bibinfo{author}{\bibfnamefont{N.~M.} \bibnamefont{Kabachnik}},
  \bibinfo{author}{\bibfnamefont{H.}~\bibnamefont{Schr{\"o}der}},
  \bibinfo{author}{\bibfnamefont{M.}~\bibnamefont{Lezius}},
  \bibnamefont{et~al.}, \bibinfo{journal}{Nature}
  \textbf{\bibinfo{volume}{446}}, \bibinfo{pages}{627} (\bibinfo{year}{2007}).

\bibitem[{\citenamefont{Uphues et~al.}(2008)\citenamefont{Uphues, Schultze,
  Kling, Uiberacker, Hendel, Heinzmann, Kabachnik, and Drescher}}]{Uphues2008}
\bibinfo{author}{\bibfnamefont{T.}~\bibnamefont{Uphues}},
  \bibinfo{author}{\bibfnamefont{M.}~\bibnamefont{Schultze}},
  \bibinfo{author}{\bibfnamefont{M.~F.} \bibnamefont{Kling}},
  \bibinfo{author}{\bibfnamefont{M.}~\bibnamefont{Uiberacker}},
  \bibinfo{author}{\bibfnamefont{S.}~\bibnamefont{Hendel}},
  \bibinfo{author}{\bibfnamefont{U.}~\bibnamefont{Heinzmann}},
  \bibinfo{author}{\bibfnamefont{N.~M.} \bibnamefont{Kabachnik}},
  \bibnamefont{and} \bibinfo{author}{\bibfnamefont{M.}~\bibnamefont{Drescher}},
  \bibinfo{journal}{New Journal of Physics} \textbf{\bibinfo{volume}{10}},
  \bibinfo{pages}{025009} (\bibinfo{year}{2008}),
  \urlprefix\url{http://stacks.iop.org/1367-2630/10/i=2/a=025009}.

\bibitem[{\citenamefont{Drescher et~al.}(2002)\citenamefont{Drescher,
  Hentschel, Kienberger, Uiberacker, Yakovlev, Scrinzi, Westerwalbesloh,
  Kleineberg, Heinzmann, and Krausz}}]{drescher2002time}
\bibinfo{author}{\bibfnamefont{M.}~\bibnamefont{Drescher}},
  \bibinfo{author}{\bibfnamefont{M.}~\bibnamefont{Hentschel}},
  \bibinfo{author}{\bibfnamefont{R.}~\bibnamefont{Kienberger}},
  \bibinfo{author}{\bibfnamefont{M.}~\bibnamefont{Uiberacker}},
  \bibinfo{author}{\bibfnamefont{V.}~\bibnamefont{Yakovlev}},
  \bibinfo{author}{\bibfnamefont{A.}~\bibnamefont{Scrinzi}},
  \bibinfo{author}{\bibfnamefont{T.}~\bibnamefont{Westerwalbesloh}},
  \bibinfo{author}{\bibfnamefont{U.}~\bibnamefont{Kleineberg}},
  \bibinfo{author}{\bibfnamefont{U.}~\bibnamefont{Heinzmann}},
  \bibnamefont{and} \bibinfo{author}{\bibfnamefont{F.}~\bibnamefont{Krausz}},
  \bibinfo{journal}{Nature} \textbf{\bibinfo{volume}{419}},
  \bibinfo{pages}{803} (\bibinfo{year}{2002}), \bibinfo{note}{article},
  \urlprefix\url{http://dx.doi.org/10.1038/nature01143}.

\bibitem[{\citenamefont{Zherebtsov et~al.}(2011)\citenamefont{Zherebtsov,
  Wirth, Uphues, Znakovskaya, Herrwerth, Gagnon, Korbman, Yakovlev, Vrakking,
  Drescher et~al.}}]{zherebtsov2011attosecond}
\bibinfo{author}{\bibfnamefont{S.}~\bibnamefont{Zherebtsov}},
  \bibinfo{author}{\bibfnamefont{A.}~\bibnamefont{Wirth}},
  \bibinfo{author}{\bibfnamefont{T.}~\bibnamefont{Uphues}},
  \bibinfo{author}{\bibfnamefont{I.}~\bibnamefont{Znakovskaya}},
  \bibinfo{author}{\bibfnamefont{O.}~\bibnamefont{Herrwerth}},
  \bibinfo{author}{\bibfnamefont{J.}~\bibnamefont{Gagnon}},
  \bibinfo{author}{\bibfnamefont{M.}~\bibnamefont{Korbman}},
  \bibinfo{author}{\bibfnamefont{V.~S.} \bibnamefont{Yakovlev}},
  \bibinfo{author}{\bibfnamefont{M.}~\bibnamefont{Vrakking}},
  \bibinfo{author}{\bibfnamefont{M.}~\bibnamefont{Drescher}},
  \bibnamefont{et~al.}, \bibinfo{journal}{Journal of Physics B: Atomic,
  Molecular and Optical Physics} \textbf{\bibinfo{volume}{44}},
  \bibinfo{pages}{105601} (\bibinfo{year}{2011}).

\bibitem[{\citenamefont{Schins et~al.}(1994)\citenamefont{Schins, Breger,
  Agostini, Constantinescu, Muller, Grillon, Antonetti, and
  Mysyrowicz}}]{Schins-PhysRevLett.73.2180}
\bibinfo{author}{\bibfnamefont{J.~M.} \bibnamefont{Schins}},
  \bibinfo{author}{\bibfnamefont{P.}~\bibnamefont{Breger}},
  \bibinfo{author}{\bibfnamefont{P.}~\bibnamefont{Agostini}},
  \bibinfo{author}{\bibfnamefont{R.~C.} \bibnamefont{Constantinescu}},
  \bibinfo{author}{\bibfnamefont{H.~G.} \bibnamefont{Muller}},
  \bibinfo{author}{\bibfnamefont{G.}~\bibnamefont{Grillon}},
  \bibinfo{author}{\bibfnamefont{A.}~\bibnamefont{Antonetti}},
  \bibnamefont{and}
  \bibinfo{author}{\bibfnamefont{A.}~\bibnamefont{Mysyrowicz}},
  \bibinfo{journal}{Phys. Rev. Lett.} \textbf{\bibinfo{volume}{73}},
  \bibinfo{pages}{2180} (\bibinfo{year}{1994}),
  \urlprefix\url{https://link.aps.org/doi/10.1103/PhysRevLett.73.2180}.

\bibitem[{\citenamefont{Kazansky et~al.}(2011)\citenamefont{Kazansky, Sazhina,
  and Kabachnik}}]{Kazansky.2011}
\bibinfo{author}{\bibfnamefont{A.~K.} \bibnamefont{Kazansky}},
  \bibinfo{author}{\bibfnamefont{I.~P.} \bibnamefont{Sazhina}},
  \bibnamefont{and} \bibinfo{author}{\bibfnamefont{N.~M.}
  \bibnamefont{Kabachnik}}, \bibinfo{journal}{Journal of Physics B: Atomic,
  Molecular and Optical Physics} \textbf{\bibinfo{volume}{44}},
  \bibinfo{pages}{215601} (\bibinfo{year}{2011}),
  \urlprefix\url{http://stacks.iop.org/0953-4075/44/i=21/a=215601}.

\bibitem[{\citenamefont{Krausz and Ivanov}(2009)}]{RevModPhys.81.163}
\bibinfo{author}{\bibfnamefont{F.}~\bibnamefont{Krausz}} \bibnamefont{and}
  \bibinfo{author}{\bibfnamefont{M.}~\bibnamefont{Ivanov}},
  \bibinfo{journal}{Rev. Mod. Phys.} \textbf{\bibinfo{volume}{81}},
  \bibinfo{pages}{163} (\bibinfo{year}{2009}),
  \urlprefix\url{https://link.aps.org/doi/10.1103/RevModPhys.81.163}.

\bibitem[{\citenamefont{Gao and Inganas}(2014)}]{GI.2014}
\bibinfo{author}{\bibfnamefont{F.}~\bibnamefont{Gao}} \bibnamefont{and}
  \bibinfo{author}{\bibfnamefont{O.}~\bibnamefont{Inganas}},
  \bibinfo{journal}{Phys. Chem. Chem. Phys.} \textbf{\bibinfo{volume}{16}},
  \bibinfo{pages}{20291} (\bibinfo{year}{2014}).

\bibitem[{\citenamefont{Song et~al.}(2016)\citenamefont{Song, Li, Ma,
  Pullerits, and Sun}}]{SLMPS.2016}
\bibinfo{author}{\bibfnamefont{P.}~\bibnamefont{Song}},
  \bibinfo{author}{\bibfnamefont{Y.}~\bibnamefont{Li}},
  \bibinfo{author}{\bibfnamefont{F.}~\bibnamefont{Ma}},
  \bibinfo{author}{\bibfnamefont{T.}~\bibnamefont{Pullerits}},
  \bibnamefont{and} \bibinfo{author}{\bibfnamefont{M.}~\bibnamefont{Sun}},
  \bibinfo{journal}{The Chemical Record} \textbf{\bibinfo{volume}{16}},
  \bibinfo{pages}{734} (\bibinfo{year}{2016}), ISSN \bibinfo{issn}{1528-0691},
  \urlprefix\url{http://dx.doi.org/10.1002/tcr.201500244}.

\bibitem[{\citenamefont{Rozzi et~al.}(2018)\citenamefont{Rozzi, Troiani, and
  Tavernelli}}]{RTT.2017}
\bibinfo{author}{\bibfnamefont{C.~A.} \bibnamefont{Rozzi}},
  \bibinfo{author}{\bibfnamefont{F.}~\bibnamefont{Troiani}}, \bibnamefont{and}
  \bibinfo{author}{\bibfnamefont{I.}~\bibnamefont{Tavernelli}},
  \bibinfo{journal}{J. Phys.: Condens. Matter} \textbf{\bibinfo{volume}{30}},
  \bibinfo{pages}{013002} (\bibinfo{year}{2018}),
  \urlprefix\url{http://stacks.iop.org/0953-8984/30/i=1/a=013002}.

\bibitem[{\citenamefont{Nisoli et~al.}(2017)\citenamefont{Nisoli, Decleva,
  Calegari, Palacios, and Mart\'in}}]{Nisoli-review}
\bibinfo{author}{\bibfnamefont{M.}~\bibnamefont{Nisoli}},
  \bibinfo{author}{\bibfnamefont{P.}~\bibnamefont{Decleva}},
  \bibinfo{author}{\bibfnamefont{F.}~\bibnamefont{Calegari}},
  \bibinfo{author}{\bibfnamefont{A.}~\bibnamefont{Palacios}}, \bibnamefont{and}
  \bibinfo{author}{\bibfnamefont{F.}~\bibnamefont{Mart\'in}},
  \bibinfo{journal}{Chemical Reviews} \textbf{\bibinfo{volume}{117}},
  \bibinfo{pages}{10760} (\bibinfo{year}{2017}),
  \urlprefix\url{http://dx.doi.org/10.1021/acs.chemrev.6b00453}.

\bibitem[{\citenamefont{Runge and Gross}(1984)}]{RungeGross:84}
\bibinfo{author}{\bibfnamefont{E.}~\bibnamefont{Runge}} \bibnamefont{and}
  \bibinfo{author}{\bibfnamefont{E.~K.~U.} \bibnamefont{Gross}},
  \bibinfo{journal}{Phys. Rev. Lett.} \textbf{\bibinfo{volume}{52}},
  \bibinfo{pages}{997} (\bibinfo{year}{1984}),
  \urlprefix\url{https://link.aps.org/doi/10.1103/PhysRevLett.52.997}.

\bibitem[{\citenamefont{Ullrich}(2012)}]{Ullrich:12}
\bibinfo{author}{\bibfnamefont{C.}~\bibnamefont{Ullrich}},
  \emph{\bibinfo{title}{Time-Dependent Density-Functional Theory}}
  (\bibinfo{publisher}{Oxford University Press}, \bibinfo{address}{Oxford},
  \bibinfo{year}{2012}).

\bibitem[{\citenamefont{Maitra}(2016)}]{Maitra.2016}
\bibinfo{author}{\bibfnamefont{N.~T.} \bibnamefont{Maitra}},
  \bibinfo{journal}{The Journal of Chemical Physics}
  \textbf{\bibinfo{volume}{144}}, \bibinfo{pages}{220901}
  (\bibinfo{year}{2016}), \urlprefix\url{https://doi.org/10.1063/1.4953039}.

\bibitem[{\citenamefont{Cucinotta et~al.}(2012)\citenamefont{Cucinotta, Hughes,
  and Ballone}}]{PhysRevB.86.045114}
\bibinfo{author}{\bibfnamefont{C.~S.} \bibnamefont{Cucinotta}},
  \bibinfo{author}{\bibfnamefont{D.}~\bibnamefont{Hughes}}, \bibnamefont{and}
  \bibinfo{author}{\bibfnamefont{P.}~\bibnamefont{Ballone}},
  \bibinfo{journal}{Phys. Rev. B} \textbf{\bibinfo{volume}{86}},
  \bibinfo{pages}{045114} (\bibinfo{year}{2012}),
  \urlprefix\url{https://link.aps.org/doi/10.1103/PhysRevB.86.045114}.

\bibitem[{\citenamefont{Covito et~al.}(2018)\citenamefont{Covito, Perfetto,
  Rubio, and Stefanucci}}]{Covito2018}
\bibinfo{author}{\bibfnamefont{F.}~\bibnamefont{Covito}},
  \bibinfo{author}{\bibfnamefont{E.}~\bibnamefont{Perfetto}},
  \bibinfo{author}{\bibfnamefont{A.}~\bibnamefont{Rubio}}, \bibnamefont{and}
  \bibinfo{author}{\bibfnamefont{G.}~\bibnamefont{Stefanucci}},
  \bibinfo{journal}{Phys. Rev. A} \textbf{\bibinfo{volume}{97}},
  \bibinfo{pages}{061401} (\bibinfo{year}{2018}),
  \urlprefix\url{https://link.aps.org/doi/10.1103/PhysRevA.97.061401}.

\bibitem[{\citenamefont{Kurth et~al.}(2005)\citenamefont{Kurth, Stefanucci,
  Almbladh, Rubio, and Gross}}]{ksarg.2005}
\bibinfo{author}{\bibfnamefont{S.}~\bibnamefont{Kurth}},
  \bibinfo{author}{\bibfnamefont{G.}~\bibnamefont{Stefanucci}},
  \bibinfo{author}{\bibfnamefont{C.-O.} \bibnamefont{Almbladh}},
  \bibinfo{author}{\bibfnamefont{A.}~\bibnamefont{Rubio}}, \bibnamefont{and}
  \bibinfo{author}{\bibfnamefont{E.~K.~U.} \bibnamefont{Gross}},
  \bibinfo{journal}{Phys. Rev. B} \textbf{\bibinfo{volume}{72}},
  \bibinfo{pages}{035308} (\bibinfo{year}{2005}),
  \urlprefix\url{https://link.aps.org/doi/10.1103/PhysRevB.72.035308}.

\bibitem[{\citenamefont{Verdozzi et~al.}(2006)\citenamefont{Verdozzi,
  Stefanucci, and Almbladh}}]{vsa.2006}
\bibinfo{author}{\bibfnamefont{C.}~\bibnamefont{Verdozzi}},
  \bibinfo{author}{\bibfnamefont{G.}~\bibnamefont{Stefanucci}},
  \bibnamefont{and} \bibinfo{author}{\bibfnamefont{C.-O.}
  \bibnamefont{Almbladh}}, \bibinfo{journal}{Phys. Rev. Lett.}
  \textbf{\bibinfo{volume}{97}}, \bibinfo{pages}{046603}
  (\bibinfo{year}{2006}),
  \urlprefix\url{https://link.aps.org/doi/10.1103/PhysRevLett.97.046603}.

\bibitem[{\citenamefont{Stefanucci et~al.}(2008)\citenamefont{Stefanucci,
  Kurth, Rubio, and Gross}}]{Stefanuccipumping}
\bibinfo{author}{\bibfnamefont{G.}~\bibnamefont{Stefanucci}},
  \bibinfo{author}{\bibfnamefont{S.}~\bibnamefont{Kurth}},
  \bibinfo{author}{\bibfnamefont{A.}~\bibnamefont{Rubio}}, \bibnamefont{and}
  \bibinfo{author}{\bibfnamefont{E.~K.~U.} \bibnamefont{Gross}},
  \bibinfo{journal}{Phys. Rev. B} \textbf{\bibinfo{volume}{77}},
  \bibinfo{pages}{075339} (\bibinfo{year}{2008}),
  \urlprefix\url{https://link.aps.org/doi/10.1103/PhysRevB.77.075339}.

\bibitem[{\citenamefont{My\"oh\"anen et~al.}(2009)\citenamefont{My\"oh\"anen,
  Stan, Stefanucci, and van Leeuwen}}]{mssvl.2009}
\bibinfo{author}{\bibfnamefont{P.}~\bibnamefont{My\"oh\"anen}},
  \bibinfo{author}{\bibfnamefont{A.}~\bibnamefont{Stan}},
  \bibinfo{author}{\bibfnamefont{G.}~\bibnamefont{Stefanucci}},
  \bibnamefont{and} \bibinfo{author}{\bibfnamefont{R.}~\bibnamefont{van
  Leeuwen}}, \bibinfo{journal}{Phys. Rev. B} \textbf{\bibinfo{volume}{80}},
  \bibinfo{pages}{115107} (\bibinfo{year}{2009}),
  \urlprefix\url{https://link.aps.org/doi/10.1103/PhysRevB.80.115107}.

\bibitem[{\citenamefont{Stefanucci et~al.}(2010)\citenamefont{Stefanucci,
  Perfetto, and Cini}}]{spc.2010}
\bibinfo{author}{\bibfnamefont{G.}~\bibnamefont{Stefanucci}},
  \bibinfo{author}{\bibfnamefont{E.}~\bibnamefont{Perfetto}}, \bibnamefont{and}
  \bibinfo{author}{\bibfnamefont{M.}~\bibnamefont{Cini}},
  \bibinfo{journal}{Phys. Rev. B} \textbf{\bibinfo{volume}{81}},
  \bibinfo{pages}{115446} (\bibinfo{year}{2010}),
  \urlprefix\url{https://link.aps.org/doi/10.1103/PhysRevB.81.115446}.

\bibitem[{\citenamefont{Hentschel et~al.}(2001)\citenamefont{Hentschel,
  Kienberger, Spielmann, Reider, Milosevic, Brabec, Corkum, Heinzmann,
  Drescher, and Krausz}}]{attometrology}
\bibinfo{author}{\bibfnamefont{M.}~\bibnamefont{Hentschel}},
  \bibinfo{author}{\bibfnamefont{R.}~\bibnamefont{Kienberger}},
  \bibinfo{author}{\bibfnamefont{C.}~\bibnamefont{Spielmann}},
  \bibinfo{author}{\bibfnamefont{G.~A.} \bibnamefont{Reider}},
  \bibinfo{author}{\bibfnamefont{N.}~\bibnamefont{Milosevic}},
  \bibinfo{author}{\bibfnamefont{T.}~\bibnamefont{Brabec}},
  \bibinfo{author}{\bibfnamefont{P.}~\bibnamefont{Corkum}},
  \bibinfo{author}{\bibfnamefont{U.}~\bibnamefont{Heinzmann}},
  \bibinfo{author}{\bibfnamefont{M.}~\bibnamefont{Drescher}}, \bibnamefont{and}
  \bibinfo{author}{\bibfnamefont{F.}~\bibnamefont{Krausz}},
  \bibinfo{journal}{Nature} \textbf{\bibinfo{volume}{414}},
  \bibinfo{pages}{509} (\bibinfo{year}{2001}),
  \urlprefix\url{http://dx.doi.org/10.1038/35107000}.

\bibitem[{\citenamefont{Lipavsk\'y et~al.}(1986)\citenamefont{Lipavsk\'y,
  \ifmmode \check{S}\else \v{S}\fi{}pi\ifmmode~\check{c}\else \v{c}\fi{}ka, and
  Velick\'y}}]{PhysRevB.34.6933}
\bibinfo{author}{\bibfnamefont{P.}~\bibnamefont{Lipavsk\'y}},
  \bibinfo{author}{\bibfnamefont{V.}~\bibnamefont{\ifmmode \check{S}\else
  \v{S}\fi{}pi\ifmmode~\check{c}\else \v{c}\fi{}ka}}, \bibnamefont{and}
  \bibinfo{author}{\bibfnamefont{B.}~\bibnamefont{Velick\'y}},
  \bibinfo{journal}{Phys. Rev. B} \textbf{\bibinfo{volume}{34}},
  \bibinfo{pages}{6933} (\bibinfo{year}{1986}),
  \urlprefix\url{https://link.aps.org/doi/10.1103/PhysRevB.34.6933}.

\bibitem[{\citenamefont{Almbladh et~al.}(1989)\citenamefont{Almbladh, Morales,
  and Grossmann}}]{PhysRevB.39.3489}
\bibinfo{author}{\bibfnamefont{C.-O.} \bibnamefont{Almbladh}},
  \bibinfo{author}{\bibfnamefont{A.~L.} \bibnamefont{Morales}},
  \bibnamefont{and}
  \bibinfo{author}{\bibfnamefont{G.}~\bibnamefont{Grossmann}},
  \bibinfo{journal}{Phys. Rev. B} \textbf{\bibinfo{volume}{39}},
  \bibinfo{pages}{3489} (\bibinfo{year}{1989}),
  \urlprefix\url{https://link.aps.org/doi/10.1103/PhysRevB.39.3489}.

\bibitem[{\citenamefont{Almbladh and Morales}(1989)}]{PhysRevB.39.3503}
\bibinfo{author}{\bibfnamefont{C.-O.} \bibnamefont{Almbladh}} \bibnamefont{and}
  \bibinfo{author}{\bibfnamefont{A.~L.} \bibnamefont{Morales}},
  \bibinfo{journal}{Phys. Rev. B} \textbf{\bibinfo{volume}{39}},
  \bibinfo{pages}{3503} (\bibinfo{year}{1989}),
  \urlprefix\url{https://link.aps.org/doi/10.1103/PhysRevB.39.3503}.

\bibitem[{\citenamefont{Perfetto and Stefanucci}(2018)}]{PS-cheers}
\bibinfo{author}{\bibfnamefont{E.}~\bibnamefont{Perfetto}} \bibnamefont{and}
  \bibinfo{author}{\bibfnamefont{G.}~\bibnamefont{Stefanucci}},
  \bibinfo{journal}{in preparation}  (\bibinfo{year}{2018}).

\bibitem[{\citenamefont{Bostr\"om et~al.}(2018)\citenamefont{Bostr\"om,
  Mikkelsen, Verdozzi, Perfetto, and Stefanucci}}]{C60paper2018}
\bibinfo{author}{\bibfnamefont{E.~V.} \bibnamefont{Bostr\"om}},
  \bibinfo{author}{\bibfnamefont{A.}~\bibnamefont{Mikkelsen}},
  \bibinfo{author}{\bibfnamefont{C.}~\bibnamefont{Verdozzi}},
  \bibinfo{author}{\bibfnamefont{E.}~\bibnamefont{Perfetto}}, \bibnamefont{and}
  \bibinfo{author}{\bibfnamefont{G.}~\bibnamefont{Stefanucci}},
  \bibinfo{journal}{Nano Lett.} \textbf{\bibinfo{volume}{18}},
  \bibinfo{pages}{785} (\bibinfo{year}{2018}),
  \urlprefix\url{https://doi.org/10.1021/acs.nanolett.7b03995}.

\bibitem[{\citenamefont{Perfetto et~al.}(2018)\citenamefont{Perfetto, Sangalli,
  Marini, and Stefanucci}}]{PSMS.2018}
\bibinfo{author}{\bibfnamefont{E.}~\bibnamefont{Perfetto}},
  \bibinfo{author}{\bibfnamefont{D.}~\bibnamefont{Sangalli}},
  \bibinfo{author}{\bibfnamefont{A.}~\bibnamefont{Marini}}, \bibnamefont{and}
  \bibinfo{author}{\bibfnamefont{G.}~\bibnamefont{Stefanucci}},
  \bibinfo{journal}{The Journal of Physical Chemistry Letters}
  \textbf{\bibinfo{volume}{9}}, \bibinfo{pages}{1353} (\bibinfo{year}{2018}),
  \urlprefix\url{https://doi.org/10.1021/acs.jpclett.8b00025}.

\bibitem[{\citenamefont{Sawatzky}(1977)}]{Sawatzky1977}
\bibinfo{author}{\bibfnamefont{G.~A.} \bibnamefont{Sawatzky}},
  \bibinfo{journal}{Phys. Rev. Lett.} \textbf{\bibinfo{volume}{39}},
  \bibinfo{pages}{504} (\bibinfo{year}{1977}),
  \urlprefix\url{https://link.aps.org/doi/10.1103/PhysRevLett.39.504}.

\bibitem[{\citenamefont{Cini}(1977)}]{CINI1977}
\bibinfo{author}{\bibfnamefont{M.}~\bibnamefont{Cini}}, \bibinfo{journal}{Solid
  State Communications} \textbf{\bibinfo{volume}{24}}, \bibinfo{pages}{681 }
  (\bibinfo{year}{1977}), ISSN \bibinfo{issn}{0038-1098},
  \urlprefix\url{http://www.sciencedirect.com/science/article/pii/0038109877903908}.

\bibitem[{\citenamefont{Sham and Schl\"uter}(1983)}]{ShamSchluter}
\bibinfo{author}{\bibfnamefont{L.~J.} \bibnamefont{Sham}} \bibnamefont{and}
  \bibinfo{author}{\bibfnamefont{M.}~\bibnamefont{Schl\"uter}},
  \bibinfo{journal}{Phys. Rev. Lett.} \textbf{\bibinfo{volume}{51}},
  \bibinfo{pages}{1888} (\bibinfo{year}{1983}),
  \urlprefix\url{https://link.aps.org/doi/10.1103/PhysRevLett.51.1888}.

\bibitem[{\citenamefont{van Leeuwen}(1996)}]{vanLeeuwenSS}
\bibinfo{author}{\bibfnamefont{R.}~\bibnamefont{van Leeuwen}},
  \bibinfo{journal}{Phys. Rev. Lett.} \textbf{\bibinfo{volume}{76}},
  \bibinfo{pages}{3610} (\bibinfo{year}{1996}),
  \urlprefix\url{https://link.aps.org/doi/10.1103/PhysRevLett.76.3610}.

\end{thebibliography}

To summarize, we have benchmarked a recently proposed NEGF 
approach~\cite{Covito2018} against 
configuration interaction calculations in a simple 1D model atom. 
With the exception of  the 
quantitative discrepancies due to the neglect of multiple 
valence-valence scatterings, good agreement is found for the
qualitative features of the Auger process. In fact, NEGF correctly 
predicts  an exponential 
law for the core-hole refilling and an asymmetric shape of the 
Auger wavepacket characterized by a long tail with superimposed ripples 
of period $T_{r}=2\p/\e_{\rm Auger}$. 
The quantitative difference is only related to the red shift of the 
energy of the Auger electron, as demonstrated by the agreement between 
NEGF and CI2B results. We point out that for the systems that we are 
interested to study in the future, i.e., organic molecules and 
molecules of biological interest, the valence-valence repulsion is less 
than 1~eV; therefore the neglect of multiple scatterings for the 
description of the internal 
dynamics  is expected to be less relevant.

The NEGF equations (\ref{CHEERSeq})
are equations of motion for the one-particle 
density matrix in the bound sector and for the occupations of the 
continuum states, {\em not} for the Green's function. Both quantities are one-time functions like the 
charge density of TDDFT $n(\blr,t)$. In particular, in a real space 
basis $\r(\blr,\blr,t)=n(\blr,t)$. Given the tight relation between 
$\r$ and $n$ it would be interesting to use the explicit form of 
the functionals $\callI[\r,f]$ and $J_{\m}[\r,f]$ as a guide to generate 
approximate xc TDDFT potentials with memory. One possibility would be to 
combine the linearized Sham-Schl\"uter equation~\cite{ShamSchluter,vanLeeuwenSS} 
with NEGF using the Generalized Kadanoff-Baym Ansatz~\cite{PhysRevB.34.6933}.

{\em Akcknowledgements}
G.S. and E.P. acknowledge EC funding through the RISE Co-ExAN (Grant
No. GA644076).
E.P. also acknowledges funding from the European Union project 
MaX Materials design at the eXascale H2020-EINFRA-2015-1, Grant
Agreement No.
676598 and Nanoscience Foundries and
Fine Analysis-Europe H2020-INFRAIA-2014-2015, Grant Agreement No.
654360.
F.C and A.R. 
acknowledge financial support from the European Research Council 
(ERC-2015-AdG-694097), Grupos Consolidados (IT578-13) and European 
Union Horizon 2020  program under Grant Agreement  676580 (NOMAD).

\end{document}